\documentclass[aps,prb,amsmath,amssymb,amsfonts,showpacs,
               superscriptaddress,tbtags,unsortedaddress,twocolumn]{revtex4}

\usepackage{bm,graphicx}

\graphicspath{{figs/}}

\begin{document}
\title
 {
  Quenched charge disorder in CuO$_2$ spin chains:
  Experimental and numerical studies
 }

\author{R.~Leidl}
\affiliation
 {
  Institut f\"ur Theoretische Physik,
  Rheinisch-Westf\"alische Technische Hochschule Aachen,
  52056 Aachen, Germany
 }

\author{R.~Klingeler}
\affiliation
 {
  Laboratoire National des Champs Magn\'{e}tiques Puls\'{e}s,
  31432 Toulouse, France
 }
\affiliation
 {
  Leibniz Institute for Solid State and Materials Research IFW Dresden,
  01171 Dresden, Germany
 }

\author{B.~B\"uchner}
\affiliation
 {
  Leibniz Institute for Solid State and Materials Research IFW Dresden,
  01171 Dresden, Germany
 }

\author{M.~Holtschneider}
\author{W.~Selke}
\affiliation
 {
  Institut f\"ur Theoretische Physik,
  Rheinisch-Westf\"alische Technische Hochschule Aachen,
  52056 Aachen, Germany
 }

\date{May 8, 2006}

\begin{abstract}
  We report on measurements of the magnetic response of the anisotropic
  CuO$_2$ spin chains in lightly hole-doped
  La$_x$(Ca,Sr)$_{14-x}$Cu$_{24}$O$_{41}$, $x\ge5$.
  The experimental data suggest that in magnetic fields $B\gtrsim 4\mathrm{T}$
  (applied along the easy axis) the system is characterized
  by short-range spin order and quasi-static (quenched) charge disorder.
  The magnetic susceptibility $\chi(B)$ shows a broad anomaly,
  which we interpret as the remnant of a spin-flop transition.
  To corroborate this idea, we present Monte Carlo simulations of a classical,
  anisotropic Heisenberg model with randomly distributed, static holes.
  Our numerical results clearly show that the spin-flop transition
  of the pure model (without holes) is destroyed and smeared out
  due to the disorder introduced by the quasi-static holes.
  Both the numerically calculated susceptibility curves $\chi(B)$
  and the temperature dependence of the position of the anomaly
  are in qualitative agreement with the experimental data.
\end{abstract}


\maketitle

\section{Introduction}
The tendency of charge carriers for self-organization seems to be an intrinsic
property of hole-doped transition metal oxides.
One remarkable example which emphasizes the interplay of charge order
and antiferromagnetism is the formation of spatial spin and charge modulations
in the high-$T_c$ cuprates.~\cite{Kivelson_etal03}
Other examples for self-organization of holes in low-dimensional magnets
include the layered nickelates\cite{Tranquada_etal94,Wochner_etal98}
and manganites\cite{Sternlieb_etal96}, and the doped CuO$_2$ spin chain
systems such as Sr$_{14-x}$Ca$_x$Cu$_{24}$O$_{41}$
and Na$_{1+x}$CuO$_{2}$.~\cite{Takigawa_etal98,Hess_etal04,
Klingeler_etal05b,Horsch_etal05}
On the other hand, these observations also suggest that quenched disorder
plays an important role in such systems.
In the case of the half-doped manganites, the quenched structural A-site
disorder was found to enhance the fluctuation of the competing
order parameters, i.e., between the charge/orbital order
and the metallic ferromagnetism.~\cite{Akahoshi_etal03}
The example of the manganites shows that, in the case of competing phases,
quenched disorder can lead to properties that are very different
from those of a slightly impure material.~\cite{Burgy_etal01}
Quenched disorder can also significantly affect the properties of hole-doped
layered cuprates.~\cite{Dagotto05}
E.g., recent numerical results suggest that disorder effects are important
to describe the underdoped regime of the layered cuprates
and the pseudogap in these compounds.~\cite{Alvarez_etal05}

In this paper, we report on experimental and numerical studies
of the magnetic response of a cuprate model system,
i.e.\ the lightly hole-doped CuO$_2$ spin chains
in La$_x$(Ca,Sr)$_{14-x}$Cu$_{24}$O$_{41}$, with $x\ge 5$.
In these compounds, two quasi-one-dimensional (1D) magnetic structures
are realized: Cu$_2$O$_3$ spin ladders and CuO$_2$ spin chains.
The former exhibit a large spin gap
of $\Delta_\mathrm{gap} \sim 400\mathrm{K}$.~\cite{Eccleston_etal98}
Hence the magnetic response at low temperature, which is the subject
of our study, is due to the chains.
The chains consist of edge-sharing CuO$_4$ plaquet\-tes containing
Cu$^{2+}$ ions with spin $S=1/2$ and non-magnetic Zhang-Rice singlets.
The concentration of holes in the spin chains amounts to less than 10\%
and the Cu spins in the hole-free chain segments form predominantly
FM fragments since the nearest-neighbor (NN) coupling is ferromagnetic.
The NN coupling is anisotropic, thereby causing an uniaxial anisotropy
perpendicular to the CuO$_4$-plaquettes, i.e.\ along the crystallographic
$b$ axis.~\cite{YushaHayn99,Kataev_etal01}
In contrast, the magnetic coupling of Cu spins via a hole is antiferromagnetic
(AFM), as is known from a comparison with the strongly (i.e., 60\%)
hole-doped spin chains of the mother compound
Sr$_{14}$Cu$_{24}$O$_{41}$.~\cite{Regnault_etal99,Ammerahl_etal00b}
Moreover, there is a finite interchain coupling causing 3D AFM spin order
below $T_N\sim 10\mathrm{K}$.~\cite{Matsuda_etal98,Ammerahl_etal00a}
In previous papers we have argued that the spin ordered phase
at zero magnetic field is presumably also characterized
by a (short-range) charge order.~\cite{SelkPokr_etal02,Kroll_etal05}
External magnetic fields of the order of a few Tesla suppress
the long-range spin order when applied along the easy axis and cause
a short-range antiferromagnetically spin ordered and charge disordered
phase.~\cite{Ammerahl_etal00a,Kroll_etal05}
In the present paper we concentrate on the properties of the intermediate
field phase at several Tesla, which is characterized by
(i) short-range AFM spin order, and (ii) quasi-static charge disorder.

\section{Motivation of the model}\label{sec_model_motiv}

\begin{figure}
  \includegraphics[width=8.6cm]{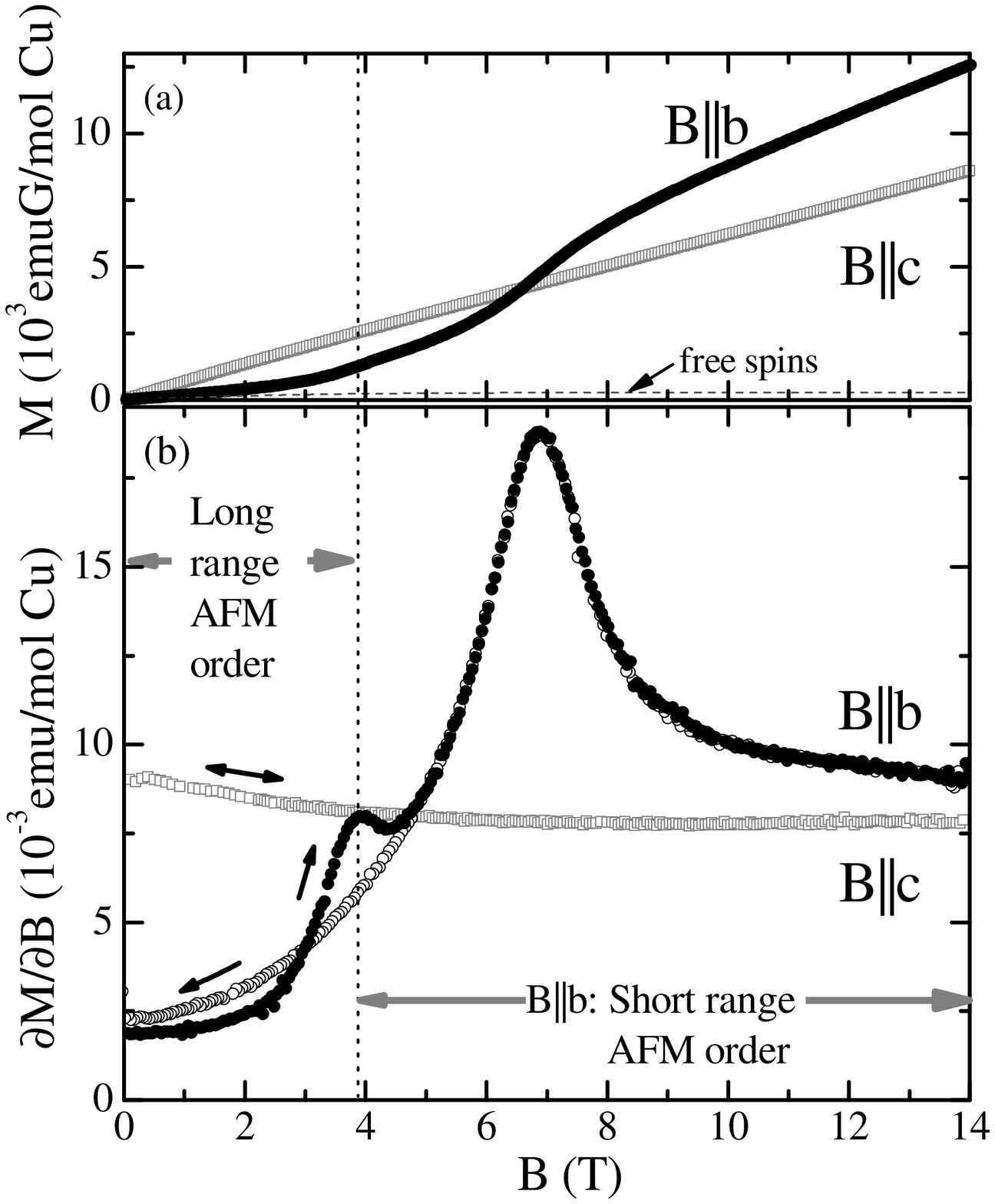}
  \caption
   {
    Magnetization $M$ (a), and susceptibility $\chi=\partial M/\partial B$ (b),
    of La$_{5.2}$Ca$_{8.8}$Cu$_{24}$O$_{41}$, at $T=2.5\mathrm{K}$,
    vs.\ magnetic field $B$ parallel to the $b$- and to the $c$-axis,
    respectively.~\cite{Kroll_etal05}
    The data are corrected by the $g$-factor taken from
    Ref.\ \onlinecite{Kataev_etal01}.
    In (a), the small, isotropic contribution due to free defect spins
    (dashed curve) has been subtracted, see Ref.\ \onlinecite{Klingeler_PhD}.
    The vertical dashed line shows the phase boundary between long-range
    and short-range antiferromagnetic spin order.
   }
  \label{fig_magnchi_la52}
\end{figure}

As was shown previously, the melting of long-range AFM spin order
at a field $B=B_1$ (depending on temperature) causes an anomaly
in the magnetization $M(B)$.~\cite{Ammerahl_etal00a,Kroll_etal05}
This is demonstrated by Fig.~\ref{fig_magnchi_la52}a, which shows
the magnetization $M(B)$, at fixed temperature $T=2.5\mathrm{K}$,
of La$_{5.2}$Ca$_{8.8}$Cu$_{24}$O$_{41}$.
If the magnetic field is oriented parallel to the chain direction,
i.e.\ $B\|c$, the magnetization depends linearly on $B$, except for a small
contribution of free spins.
In contrast, two anomalies are observed in $M(B\|b)$, which become clearly
visible if the susceptibility $\chi=\partial M/\partial B$
in Fig.\ \ref{fig_magnchi_la52}b is considered.
At $B_1=3.75\mathrm{T}$ one recognizes a sharp peak which is attributed
to the melting of the long-range spin order.
Hence, this anomaly signals a transition from a spin
and (probably short-ranged) charge ordered phase for low fields $B\|b<B_1$
into a charge disordered state for $B>B_1$.

Based on the suggestions described in Ref.\ \onlinecite{Klingeler_PhD},
various theoretical studies have been devoted to the phenomena at
$B=B_1$.~\cite{SelkPokr_etal02,HoltSelk03,HSL05,SHL05}
There the magnetic degrees of freedom were described by Ising spins
and the holes were assumed to move either freely along the chains
or under the influence of a periodic pinning potential stabilizing
a striped structure.
These models predict a breakdown of the striped (charge ordered) phase
and may thus explain the transition at $B=B_1$.

Our present study, however, focuses on the properties for $B>B_1$.
Previous numerical and experimental
work\cite{Klingeler_PhD,SelkPokr_etal02,Kroll_etal05}
implies that this phase is characterized by short-range AFM spin correlations,
and quasi-static (quenched) charge disorder.
The data in Fig.~\ref{fig_magnchi_la52}b display an additional broad peak
in $\chi$, at $B_2=6.9\mathrm{T}>B_1$, which was not captured
by the previous theoretical studies.
We attribute this anomaly to the reorientation of the Cu spins.
The idea that the anomaly at $B_2$ is in fact a ``smoothened out''
spin-flop transition will be explored in greater detail in the next section,
where we present the results of our Monte Carlo simulations.
We propose that the anomaly at $B=B_2$ is the relic of a spin-flop transition,
which is smeared out due to the strong disorder induced
by the quasi-static holes.
In this sense one may call the anomaly a ``pseudo'' spin-flop peak.

In the scenario of the ``pseudo'' spin-flop transition, the magnetic field
overcomes, for $B>B_2$, the uniaxial anisotropy which is due to the nearly
90$^\circ$ Cu-O-Cu exchange
(cf.\ Refs.\ \onlinecite{YushaHayn99,Kataev_etal01}).
Quantitatively, the experimental value of $B_2$ is consistent with a recent
inelastic neutron scattering study
on La$_5$Ca$_9$Cu$_{24}$O$_{41}$,\cite{Matsuda_etal03}
which reported a spin gap of $\Delta_\mathrm{gap}/(g\mu_B)
= (7\pm0.5)\mathrm{T}$.

\section{Numerical simulations\label{NumSim}}

\subsection{Definition of the model, choice of interaction parameters,
            and simulation method}

Taking the scenario of quenched charge disorder and short-range AFM spin order
for $B>B_1$ as our starting point, we adopt a complementary view
to the previous studies which considered mobile charge carriers
and study the influence of \emph{quenched} charge disorder on the magnetic
properties of the system, ignoring the mobility of the holes altogether.

We consider a $L\times L$ square lattice consisting of $L$ rows,
which we identify with the chains, and $L$ sites per chain.
This choice of lattice geometry is motivated by neutron scattering
experiments indicating that the copper ions in the CuO$_2$ planes
of La$_5$Ca$_{14}$Cu$_{24}$O$_{41}$ form a rectangular array.~\cite{Guk_unpub}
We conveniently set the lattice constants along and perpendicular
to the chains equal to unity.
Periodic boundary conditions are employed throughout.

Each site $(i,j)$, where $i$ is the chain index and $j$ labels the sites
along the chain, is either occupied by a spin (representing a magnetic
Cu$^{2+}$ ion), or a non-magnetic hole (Zhang-Rice singlet).
To describe the hole distribution, we introduce random variables $p_{i,j}$
taking the values $p_{i,j}=1$ if a spin resides at lattice site $(i,j)$
and $p_{i,j}=0$ if it is occupied by a hole.
The spins are modeled by (classical) three-component vectors
$\vec{S}_{i,j}=(S_{i,j}^x,S_{i,j}^y,S_{i,j}^z)$ with $|\vec{S}_{i,j}|=1$.
As discussed in Ref.\ \onlinecite{LS04b}, we expect our results to remain
qualitatively correct if one took the quantum character of the spins properly
into account (although there would be, of course, quantitative deviations).
This is basically a consequence of the Ising-like anisotropy of the model
which tends to suppress quantum fluctuations, in particular in the presence
of a field applied along the easy axis as in our case.
If a hole is at site $(i,j)$, we set $\vec{S}_{i,j}=0$.
We simulate either the pure system without holes ($p_{i,j}=1$ for all $i,j$),
or employ a fixed hole concentration of 10\% within each chain.
The latter should resemble the situation in the lightly hole-doped
chain systems of La$_x$Ca$_{14-x}$Cu$_{24}$O$_{41}$ with $x\sim5$.
Moreover, we disallow nearest-neighbor pairs of holes within the same chain,
since such configurations are energetically unfavorable due to the strong
Coulomb repulsion.
Thus consecutive holes along the chains are always separated
by at least one spin.

\begin{figure}[ht]
  \includegraphics[width=8.6cm]{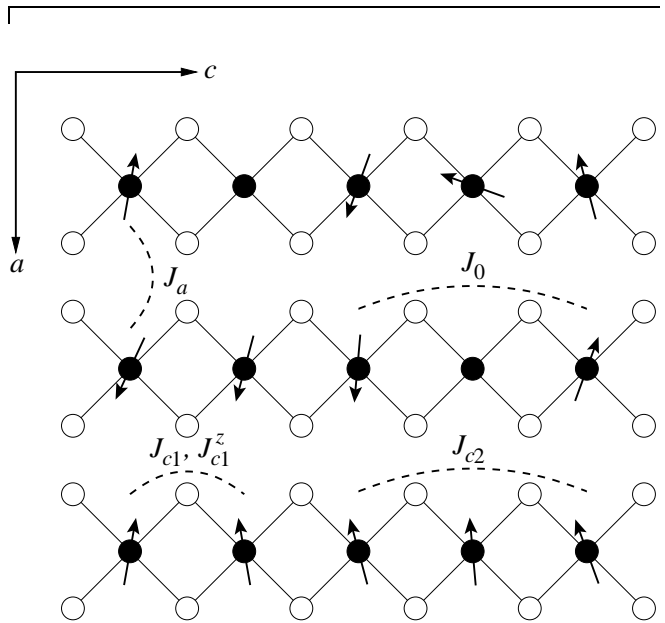}
  \caption
   {
    Schematic representation of the interactions of our model Hamiltonian,
    Eq.\ (\ref{Ham}).
    Full and open circles denote Cu and O atoms, respectively.
    Nearest-neighbor (NN) spins along the CuO$_2$ chains ($c$ direction)
    interact via an anisotropic ferromagnetic exchange ($J_{c1}$,
    $J_{c1}^z=(1+\Delta)J_{c1}$), whereas next-nearest-neighbor (NNN) spins
    are coupled antiferromagnetically.
    The strength of the antiferromagnetic coupling depends on whether
    the NNN spins are separated by a hole ($J_0$) or by another spin
    ($J_{c2}$).
    Finally, neighboring chains have an AFM interaction ($J_a$).
   }
  \label{fig_couplings}
\end{figure}

The configurational energy depends both on the spin variables
$\{\vec{S}_{i,j}\}$ and the hole distribution described by the occupation
variables $\{p_{i,j}\}$.
In a field $H$ applied along the $z$-axis, the Hamiltonian of our model reads:
\begin{widetext}
  \begin{equation}\label{Ham}
    \begin{split}
      \mathcal{H} =
       & - J_{c1}\sum_{i,j}
           \left(
            \vec{S}_{i,j}\cdot\vec{S}_{i,j+1} + \Delta\,S_{i,j}^zS_{i,j+1}^z
           \right)
         - J_{c2}\sum_{i,j}\vec{S}_{i,j}\cdot\vec{S}_{i,j+2}
         - J_0\sum_{i,j}(1-p_{i,j+1})\vec{S}_{i,j}\cdot\vec{S}_{i,j+2}\\
       & - J_a\sum_{i,j}\vec{S}_{i,j}\cdot\vec{S}_{i+1,j}
         - H\sum_{i,j}S_{i,j}^z\,.
    \end{split}
  \end{equation}
\end{widetext}
The interactions of this model are shown schematically
in Fig.\ \ref{fig_couplings}.
The Cu-O-Cu bonding angle of nearly 90$^\circ$ suggests
that the nearest-neighbor (NN) intrachain coupling $J_{c1}$
is ferromagnetic ($J_{c1}>0$).
Moreover, this coupling is
anisotropic,\cite{YushaHayn99,Ammerahl_etal00a,Kataev_etal01}
favoring the alignment of the spins along an easy axis
(the crystallographic $b$-axis), which we take to be the $z$-axis.
The anisotropy parameter is $\Delta>0$, where $\Delta=0$ corresponds
to the isotropic case.
In CuO$_2$ spin chains one expects next-nearest-neighbor (NNN) spins
to be coupled antiferromagnetically.~\cite{Mizuno_etal98}
The coupling between NNN spins is $J_{c2}<0$ if they are separated by a spin
and $J_0<0$ if a hole resides between them, where $|J_0|>|J_{c2}|$.
The difference is mainly caused by the smaller size of the Cu$^{3+}$ ions
leading to a stronger overlap of the $p$-orbitals of the involved oxygen ions.
Finally, in accordance with the experimentally observed long-range
AFM ordering, we assume an interchain coupling $J_a<0$ between NN spins
on adjacent chains.

In the following, we turn to the values of the interaction parameters
of Eq.\ (\ref{Ham}).
To the best of our knowledge, no direct information on the magnetic coupling
parameters is available which agrees with our thermodynamic
studies.~\cite{LS04a}
Thus, for the numerical simulations we indirectly estimate the interaction
parameters.
First, we address the NNN interaction across a hole.
This parameter has been determined to be $J_0/k_B=-130\mathrm{K}$
from neutron diffraction studies of the stoichiometric compound
Sr$_{14}$Cu$_{24}$O$_{41}$.~\cite{Regnault_etal99,Matsuda_etal99}
In contrast to $J_0$, only little is known about $J_{c1}$ and $J_{c2}$.
Qualitatively, the bonding geometry suggests $J_{c1}>0$
and $J_{c2}<0$.~\cite{YushaHayn99,Drechsler_etal05,Klingeler_etal06}
In addition, the presence of ferromagnetic spin order along the chains
in the weakly hole-doped compounds implies $|J_{c1}|>|J_{c2}|$.
In order to get a quantitative estimation, we apply the exchange parameters
which have been extracted for the undoped CuO$_2$ spin chains
in Li$_2$CuO$_2$.
In this compound, the Cu-O-Cu bonding angles in the CuO$_2$ spin chains
are very similar to those of (Ca,Sr)$_{14-x}$La$_x$Cu$_{24}$O$_{41}$.
Thus, we take $J_{c1}/k_B
= 100\mathrm{K}$.~\cite{Mizuno_etal99,Graaf_etal02,Klingeler_etal06}
Moreover, we assume $J_{c2}/k_B=-35\mathrm{K}$, which is again estimated
from a comparison with Li$_2$CuO$_2$ by applying the bond-valence sum rule
and the pressure dependence of $J_0$ in Sr$_{14}$Cu$_{24}$O$_{41}$.
For Sr$_{14}$Cu$_{24}$O$_{41}$, the latter amounts to
$\partial J_0/\partial p_c = 4.2\mathrm{K/GPa}$.~\cite{Ammerahl_etal00b}

The magnitude of $J_a$ significantly affects the zero-field ordering
temperature of the model (without holes).
In Sr$_{14}$Cu$_{24}$O$_{41}$, the coupling between Cu spins in adjacent
chains amounts to $J_a\simeq-20\mathrm{K}$.~\cite{Regnault_etal99}
However, preliminary diffraction experiments \cite{Guk_unpub} suggest changes
of the relative positions of neighboring CuO$_2$ chains upon La-doping,
which are supposed to strongly affect the interchain coupling constant.
According to a recent result on undoped spin chains
in Ref.\ \onlinecite{Drechsler_etal05},
\begin{equation}
  \Theta_\mathrm{CW}^\mathrm{3D} \approx
    \Theta_\mathrm{CW}^\mathrm{1D} - z_\mathrm{eff}\frac{J_a}{4},
\end{equation}
where the 3D Curie-Weiss temperature may be estimated
as $\Theta_\mathrm{CW}^\mathrm{3D} \simeq -8\mathrm{K}$ from a fit
to high-temperature susceptibility data,\cite{Klingeler_PhD}
while for the 1D Curie-Weiss temperature one has
$\Theta_\mathrm{CW}^\mathrm{1D} \simeq 0.23J_{c1} = -23\mathrm{K}$
from a cluster calculation.~\cite{Drechsler_etal05}
Using the (approximate) effective number $z_\mathrm{eff}=2$
of nearest neighbors at surrounding chains, one gets a slightly larger value
$J_a\simeq -30\mathrm{K}$ as compared to Sr$_{14}$Cu$_{24}$O$_{41}$.
This estimate, however, intimately depends on $z_\mathrm{eff}$,
which might be different.
In the following, we set $J_a/k_B = -25\mathrm{K}$, since this gives rise
to a zero-field ordering temperature which appears to be quite reasonable
as compared to the experiments (see the discussion in Sec.\ \ref{sec_pure}).

From the interaction parameters $J_{c1}$, $J_{c2}$, and $J_a$, and a fit
to the experimentally determined spin-wave gap
of Ref.\ \onlinecite{Matsuda_etal03}, we can calculate the anisotropy parameter
$\Delta$.
This yields $\Delta=0.0255$.

The model is simulated employing a single-spin Met\-ropolis algorithm.
System sizes range from $L=20$ to $L=240$.
To obtain good equilibrium data, up to $2\times10^7$ Monte Carlo steps
per site are needed for the largest systems.
At the beginning of each run, 20\% of the steps are discarded
for thermalization.
For the system with holes, we average over up to 300 randomly generated
realizations of the disorder.

A quantity of primary concern due to its relation to the experiments
is the magnetic susceptibility $\chi^z$,
\begin{equation}\label{chiz}
  \chi^z = \frac{1}{k_BT L^2}\left(\left\langle(M^z)^2\right\rangle
            - \left\langle M^z\right\rangle^2\right),
\end{equation}
where $\langle\ldots\rangle$ denotes the thermal average
and $M^z=\sum_{i,j}S_{i,j}^z$ is the $z$-component of the total magnetization.
Other observables of interest include the specific heat
and the staggered magnetization.
We also record typical spin configurations generated during
the Monte Carlo runs in order to monitor directly microscopic properties
of the system.

\subsection{Pure (undoped) system}\label{sec_pure}

Before studying the influence of the disorder, it is instructive
to review some basic properties of the pure system without holes.
In the present context, the most relevant features of the pure model are
(i) the existence, for low temperatures and fields, of a phase
with long-range AFM order perpendicular to the chains,
as well as (ii) the occurrence of a spin-flop transition upon applying
a magnetic field along the easy axis (i.e., the $z$-axis).

Let us first discuss the model in zero magnetic field ($H=0$).
To measure the long-range AFM order perpendicular to the chains,
we define the quantity
\begin{equation}\label{interchain_OP}
  M_\mathrm{s}^2 = \frac{1}{L}
   \sum_{j=1}^L\left(\frac{1}{L}\sum_{i=1}^L(-1)^i\vec{S}_{i,j}\right)^2.
\end{equation}
Note that the expression within the parentheses is the staggered
magnetization (per spin) of one column of the square lattice.
We cannot simply take the difference between the total magnetizations of even
and odd rows, which would be a natural candidate for the AFM order parameter,
since the usual AFM structure is modified by a helical ordering of the spins
along the chains, as explained below.
Thus the total magnetization of each chain vanishes for $H=0$.
In the fully (antiferromagnetically) ordered state, one has $M_\mathrm{s}^2=1$.

\begin{figure}[ht]
  \includegraphics[width=8.6cm]{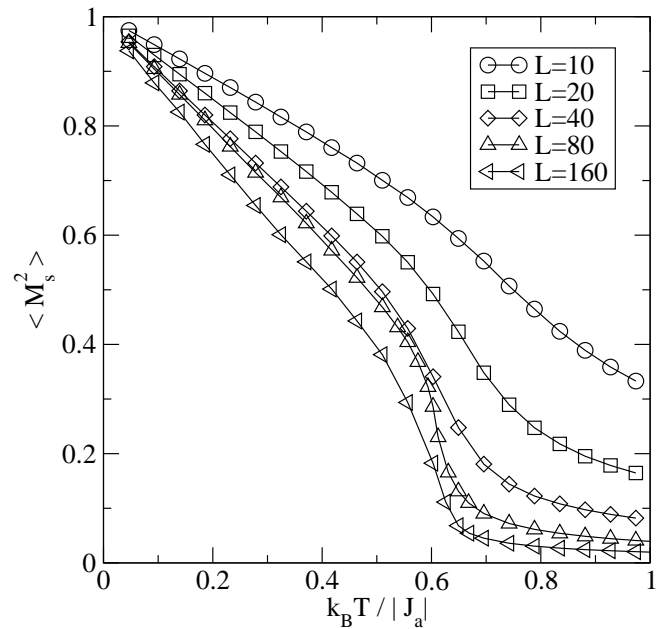}
  \caption
   {
    Interchain order parameter, Eq.\ (\ref{interchain_OP}), of the pure model
    vs.\ temperature (at $H=0$), for different system sizes $L$.
   }
  \label{fig_mstaggsqu_pure}
\end{figure}

From our simulational data (Fig.\ \ref{fig_mstaggsqu_pure}) we infer
that a phase with long-range AFM order exists at low temperatures.
The interchain order parameter $M_\mathrm{s}^2$ seems to vanish continuously
at a N\'{e}el temperature $T_N$, which we estimate as $k_BT_N/|J_a|\simeq0.61$
by finite-size extrapolation of our data.
This value is also obtained by analyzing the peak positions
of the specific heat.

Taking the spin $S=1/2$ of the Cu$^{2+}$ ions into account we obtain
an estimate of $T_N=13\mathrm{K}$ for the N\'{e}el temperature at zero field,
which is reasonably close to the experimental value
for La$_5$Ca$_9$Cu$_{24}$O$_{41}$ ($T_N=10.5\mathrm{K}$).~\cite{Matsuda_etal98}
However, this result has to be taken with care and should only be regarded
as a rough consistency check.
To mention just two points, the real system is not hole-free
and quantum fluctuations (absent in our classical spin model)
certainly alter the value of the ordering temperature
(cf.\ Ref.\ \onlinecite{LS04b}).

\begin{figure}[ht]
  \includegraphics[width=8.6cm]{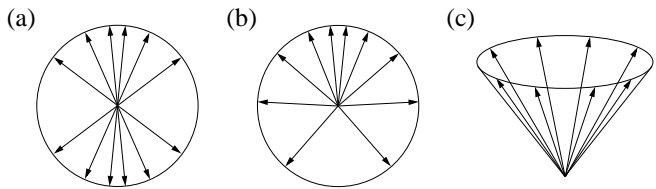}
  \caption
   {
    Schematic representation of the (intrachain) spin configurations
    for (a) $H=0$, (b) $0<H<H_\mathrm{sf}$,
    and (c) $H_\mathrm{sf}<H<H_\mathrm{pm}$, where $H_\mathrm{sf}$
    is the spin-flop field and $H_\mathrm{pm}$ the field of the spin-flop
    to paramagnetic transition.
   }
  \label{fig_spinconfigs}
\end{figure}

Whereas neighboring spins on adjacent chains are aligned antiferromagnetically
for $T<T_N$, the spins within the chains exhibit a more complicated structure
due to the competing intrachain interactions $J_{c1}>0$ and $J_{c2}<0$.
At $T=0$, this structure can be found by a ground-state analysis
using (and slightly generalizing) the methods described
in Refs.\ \onlinecite{Nagamiya_etal62} and \onlinecite{RobinErd_70}.
Without anisotropy ($\Delta=0$), one would obtain a simple helical ordering
within each chain.
In that case the spins rotate, with a constant angle $\alpha$ between
two consecutive spins, within a plane whose orientation is fixed in space.
A straightforward calculation yields $\alpha\simeq 44^\circ$, corresponding
to a wavelength of the helix of approximately eight lattice constants.
The finite exchange anisotropy $\Delta>0$, however, modifies this structure.
In order to minimize the anisotropy energy, the spins rotate in a plane
that contains the $z$-axis (without anisotropy the orientation of the plane
is arbitrary).
Moreover, the rotation angle is not a constant, but somewhat smaller for spins
in the vicinity of the $z$-axis.
The wavelength of the modified helix, though, changes only little as compared
to the isotropic case.
Such a configuration is depicted schematically in Fig.\ \ref{fig_spinconfigs}a.
The results of the ground-state analysis are corroborated by inspection
of typical low-temperature Monte Carlo configurations.
We can unambiguously identify the type of helical order
shown in Fig.\ \ref{fig_spinconfigs}a.
The wavelength of the helix turns out to depend only weakly on temperature.

A magnetic field $H>0$ along the easy axis lifts the symmetry
between the positive and negative $z$-direction
and the system develops a finite total magnetization along the $z$-axis.
At $T=0$, one can again find the corresponding spin configurations
by a ground-state analysis.
For small fields $H<H_\mathrm{sf}$, where $H_\mathrm{sf}$ is the spin-flop
field (see below), the analysis yields a fan-like structure
(Fig.\ \ref{fig_spinconfigs}b).
At $H=H_\mathrm{sf}$, this structure becomes unstable against
a spin-flop phase where the spins make a finite angle with the $z$ axis
and rotate on the surface of a cone (see Fig.\ \ref{fig_spinconfigs}c).
All spins now have the same $z$-component, but the $x$- and $y$-components
are reversed for nearest-neighbor spins on adjacent chains.
This justifies calling the structure a ``spin-flop phase''.
At $H=H_\mathrm{sf}$, the $z$-component of the magnetization (and various
other quantities) exhibit a discontinuity.
For the parameters of our model, the value of the zero-temperature
spin-flop field is given by $H_\mathrm{sf}/|J_a|\simeq 0.70$.
Upon further increasing the field the opening angle of the cone continuously
shrinks to zero until at $H=H_\mathrm{pm}$ all spins point along the $z$-axis.
However, this transition from the spin-flop to the paramagnetic phase
occurs at values of the magnetic field much larger than the highest fields
used in the experiments and will therefore be disregarded in the following.

\begin{figure}[ht]
  \includegraphics[width=8.6cm]{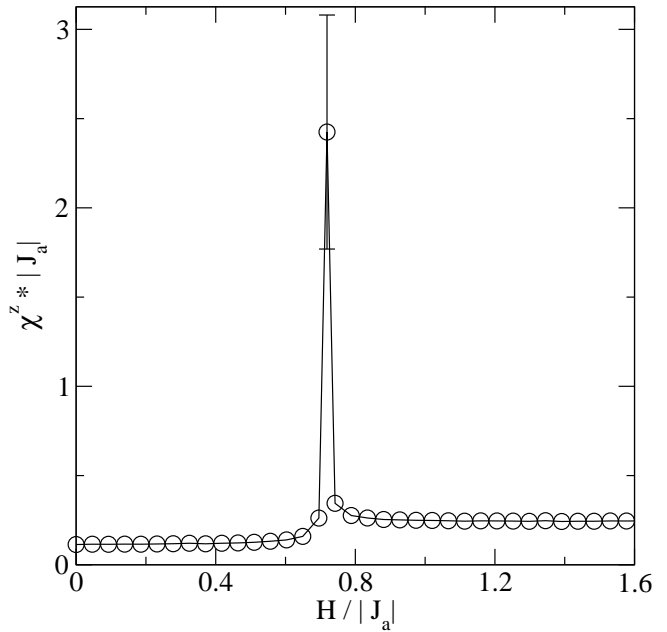}
  \caption
   {
    Magnetic susceptibility $\chi^z$ of the pure model at fixed temperature
    $k_BT/|J_a|=0.25$ for a system of size $L=80$.
    Note the sharp spin-flop peak at $H_\mathrm{sf}/|J_a|\simeq0.72$.
   }
  \label{fig_chipure_T025}
\end{figure}

The above spin structures in a magnetic field $H>0$ can again be found
in our finite-temperature Monte Carlo configurations.
Moreover, we observe a sharp peak in the susceptibility $\chi^z$
(Fig.\ \ref{fig_chipure_T025}), which occurs at a field quite close
to the value of the spin-flop field at $T=0$, $H_\mathrm{sf}/|J_a|=0.70$
(see above).
Apparently, the peak signals the (presumably first-order) transition
towards the spin-flop phase.
Similar anomalies are found in other quantities.
The spin-flop field $H_\mathrm{sf}$ is only weakly temperature-dependent
(for low temperatures).

A more detailed investigation of the phase diagram of the pure model,
the nature of the various transition lines, and the behavior near possible
critical and multicritical points, where the different phases eventually meet,
would certainly be of interest by its own but is beyond the scope
of the present paper.

\subsection{Influence of random, immobile holes}

The introduction of randomly distributed, immobile holes has a drastic impact
on the properties of the model.
Any long-range spin order (including the AFM ordering perpendicular
to the chains) gets destroyed, which leads to a smearing out
of all phase transitions discussed in the previous section.

The loss of long-range order already appears in the ground-state ($T=0$).
Within a modeling in terms of Ising rather than Heisenberg spins,
it can be shown analytically that at $T=0$ the spin correlation function
within the chains decays exponentially
for large distances.~\cite{SelkPokr_etal02}
This is intuitively clear since the strong AFM coupling $J_0$ enforces
an antiparallel alignment of two spins on the left and right sides of a hole.
Thus the chain splits up into (ferromagnetic) fragments
separated by antiphase boundaries.
If the holes are distributed randomly, all long-range spin correlations
along the chain are thus destroyed.
For the spin correlations perpendicular to the chains an analytic treatment
is much more complicated due to the frustration of the interchain interactions.
The latter occurs since fragments of neighboring chains will generally
be displaced against each other.
In order to minimize its total energy the system will thus form additional
antiphase boundaries within the chains in order to balance the competing
intrachain and interchain interaction energies.
In any event, one again expects an exponential decay of the spin correlations.
This has been confirmed numerically.

\begin{figure}[ht]
  \includegraphics[width=8.6cm]{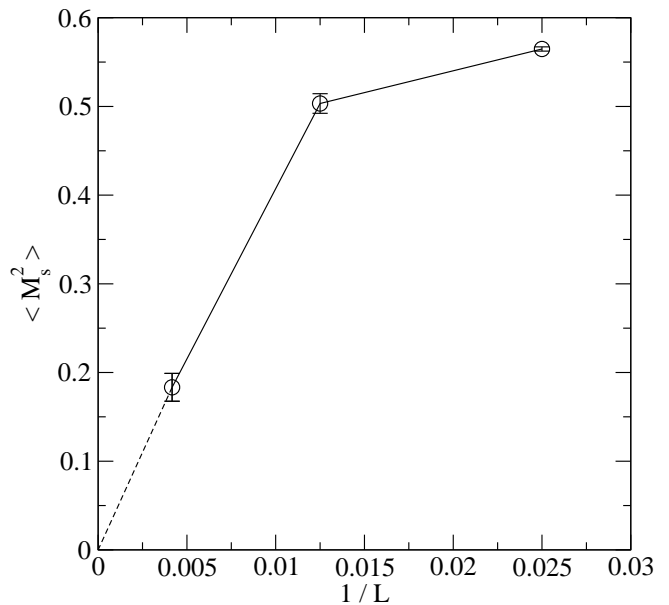}
  \caption
   {
    Interchain order parameter, Eq.\ (\ref{interchain_OP}), for various system
    sizes ($L=40,80,240$) at fixed temperature $k_BT/|J_a|=0.1$.
    For $1/L\to 0$ the order parameter seems to extrapolate to zero
    (dashed line), indicating the destruction of the long-range
    AFM order due to the quenched holes.
   }
  \label{fig_mstaggsqu_T01}
\end{figure}

The above mechanism for the destruction of long-range correlations may equally
well apply if the Ising spins are replaced by Heisenberg spins.
We have checked this by analyzing low-temperature Monte Carlo data
of the AFM interchain order parameter, Eq.\ (\ref{interchain_OP}),
for varying system sizes $L$ (Fig.\ \ref{fig_mstaggsqu_T01}).
The order parameter seems to extrapolate to zero for $L\to\infty$,
indicating the absence of long-range AFM interchain order
in the thermodynamic limit, as expected from the above arguments.

One should keep in mind that despite the lack of long-range order,
the spins will in general still exhibit some degree of short-range ordering,
which reflects the properties of the corresponding pure phases (without holes)
in the various regions of the $T,H$-plane.

The smearing out of the phase transitions due to the presence of the randomly
distributed holes can be inferred from our simulational data.
E.g., for $H=0$ the specific heat of the pure model exhibits a peak
whose height increases with the system size $L$ and whose position approaches
the N\'{e}el temperature $T_N$ as $L\to\infty$.
For the disordered system, on the other hand, we observe a non-critical
maximum, being almost size-independent for sufficiently large systems,
which occurs at a ``pseudo'' N\'{e}el temperature
$k_BT_N^{\mathrm{ps}}/|J_a|\simeq 0.58$, as compared
to $k_BT_N/|J_a|\simeq0.61$ for the pure model (see Sec.\ \ref{sec_pure}).
Thus the N\'{e}el transition is not only smeared out but also slightly shifted
towards lower temperature.
In addition, the specific heat shows a small anomaly at lower temperatures,
which is probably due to the incommensurability of the wavelength
of the helical structures with the system size.
The same conclusions are found by analyzing other quantities,
such as the magnetic susceptibility.

\begin{figure}[ht]
  \includegraphics[width=8.6cm]{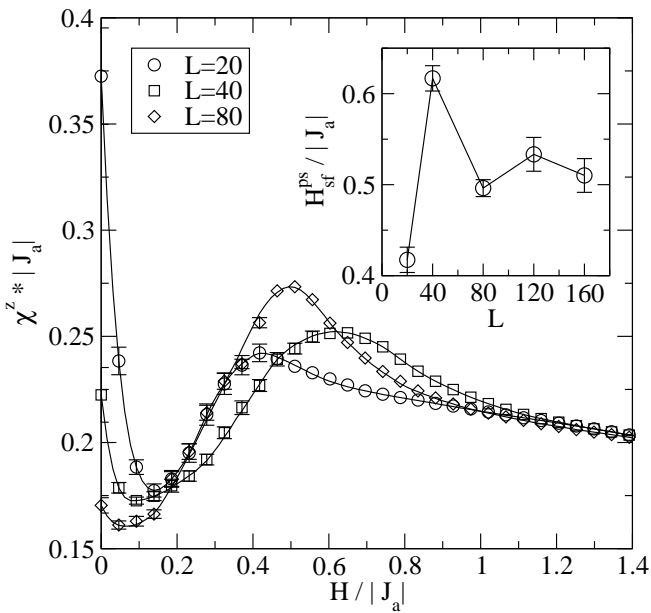}
  \caption
   {
    Susceptibility $\chi^z$ vs.\ magnetic field at constant temperature
    $k_BT/|J_a|=0.25$.
    The solid lines are guides to the eye.
    Up to 300 disorder realizations were used to generate the data.
    The error bars are smaller than the symbol sizes to the right
    of the maxima and are thus not shown there.
    The inset shows the position of the maximum as a function of system size.
   }
  \label{fig_chi_T025}
\end{figure}

Turning now to the (in our context) more interesting case of a non-vanishing
magnetic field $H\neq0$, we examine how the spin-flop transition
of the pure system is affected by the quenched holes.
Again, we find that the transition is transformed into a smooth anomaly.
Whereas for the pure system the susceptibility $\chi^z$ exhibits a sharp peak
at $H=H_\mathrm{sf}$ (cf.\ Fig.\ \ref{fig_chipure_T025}),
we now observe a broad (and much smaller) maximum
at a ``pseudo'' spin-flop field $H_\mathrm{sf}^\mathrm{ps}$
(depending on system size), see Fig.\ \ref{fig_chi_T025}.
The curves display some finite-size dependence for small systems.
E.g., the strong increase, for small system sizes, of $\chi^z$ as $H\to0$
is weakened significantly for larger systems.
For systems larger than $L=80$ the curves change only little.
The inset of Fig.\ \ref{fig_chi_T025} illustrates this for the position
of the maximum, which approaches $H_\mathrm{sf}^\mathrm{ps}/|J_a| \approx0.5$.
Analogous conclusions apply to the height of the maximum,
which quickly saturates if $L\gtrsim80$.
Thus, in the disordered case, it seems to be sufficient to simulate systems
of size $L=80$ to capture the relevant properties holding
in the thermodynamic limit.

A closer look at the Monte Carlo configurations reveals local spin-flop
structures for $H\gtrsim H_\mathrm{sf}^\mathrm{ps}$,
while for $H\lesssim H_\mathrm{sf}^\mathrm{ps}$ domains
showing helically modified antiferromagnetic structures can be identified.
In this sense, $H_\mathrm{sf}^\mathrm{ps}$ marks a smooth crossover
from the AFM phase to the spin-flop state.
This qualitative picture can be corroborated by examining
a suitable quantity measuring the local AFM order (see below).

For magnetic fields smaller than the spin-flop field
$H\lesssim H_\mathrm{sf}^\mathrm{ps}$, the disorder fluctuations
due to different realizations of the hole distribution are significantly
larger than for $H \gtrsim H_\mathrm{sf}^\mathrm{ps}$.
To reduce the disorder fluctuations one therefore has to average over many
realizations.
Since simulating many different hole distributions for all values of the field
requires too much computational time, we generated a large number of up
to 300 disorder realizations for lower fields only
($H < H_\mathrm{sf}^\mathrm{ps}$).
For higher fields, 100 realizations usually turned out to be sufficient.
In this way we obtained a reasonably good statistics for all data points.

\begin{figure}[ht]
  \includegraphics[width=8.6cm]{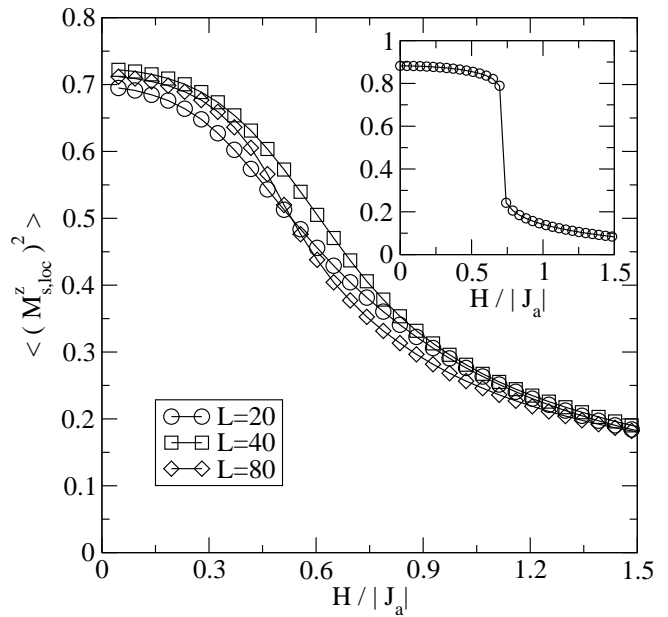}
  \caption
   {
    Square of the $z$-component of the local staggered magnetization,
    Eq.\ (\ref{mzstaggsqu}), vs.\ magnetic field at constant temperature
    $k_BT/|J_a|=0.25$, indicating a smooth transition between the AFM
    and the spin-flop phase.
    For the pure model, the same quantity appears to behave discontinuously
    at the spin-flop field (see inset).
    In all cases, the error bars are much smaller than the symbol size.
   }
  \label{fig_mzstaggsqu}
\end{figure}
To provide further evidence that the broad susceptibility maximum is indeed
the remnant of a smeared-out spin-flop transition, we study the square
of the $z$-component of the local staggered magnetization, which is a measure
of the degree of local AFM order perpendicular to the chains:
\begin{equation}\label{mzstaggsqu}
  (M_\mathrm{s,loc}^z)^2 = \frac{1}{4L^2}\sum_{i=1}^L\sum_{j=1}^L
   \left[S^z_{i,j}-(S^z_{i-1,j}+S^z_{i+1,j})/2\right]^2.
\end{equation}
Note that the expression under the double sum is (up to a factor of $1/4$)
the square of the $z$-component of the local AFM order parameter
at site $(i,j)$.
This local quantity is then averaged over the whole lattice.
As exemplified in Fig.\ \ref{fig_mzstaggsqu},
$\langle(M_\mathrm{s,loc}^z)^2\rangle$ drops down smoothly as one increases
the magnetic field, i.e.\ the local AFM order along the $z$-direction
decreases, as one expects for a transition between an AFM
and a spin-flop phase.
Moreover, the slope of the curve is maximal at the same field
$H_\mathrm{sf}^\mathrm{ps}$ where the susceptibility $\chi^z$ has its maximum
(cf.\ Fig.\ \ref{fig_chi_T025}).
Note that $H_\mathrm{sf}^\mathrm{ps}$ is somewhat lower
than the spin-flop field $H_\mathrm{sf}/|J_a|\simeq0.70$ of the pure model.
For the pure system, $\langle(M_\mathrm{s,loc}^z)^2\rangle$ appears to jump
at $H=H_\mathrm{sf}$ (see the inset of Fig.\ \ref{fig_mzstaggsqu}).

We would like to stress that many of the above conclusions
are qualitatively insensitive to details of the model like the precise values
of the interaction parameters, provided that the system exhibits
randomly distributed, immobile holes.
In fact, we have also carried out simulations using the model proposed
in Ref.\ \onlinecite{Matsuda_etal03}, which was based on an interpretation
of inelastic neutron scattering data for La$_5$Ca$_9$Cu$_{24}$O$_{41}$
(an analysis of this  model, with and without mobile holes,
may be found in Ref.\ \onlinecite{LS04b}).
This model has a different lattice geometry, a single-ion instead of an
exchange anisotropy and quite distinct values of the interaction parameters.
Moreover, we considered a simplified model with vanishing NNN coupling
within the chains and a ferromagnetic NN interaction
(i.e., $J_{c2}=0$ and $J_{c1}>0$).
Essentially all of our conclusions concerning the smearing out
of the phase transitions, which transform into (smooth) anomalies
when introducing quenched holes, also hold for these modified models.
On the other hand, when comparing the results of the simulations
with the experimental data, the quantitative agreement seems to be
most satisfying for our present model.

However, the choice of the interaction parameters certainly has an effect
on the typical spin configurations within the chains.
E.g., for the simplified model with $J_{c2}=0, J_{c1}>0$ mentioned above
the chains order ferromagnetically (this also applies to the model
of Ref.\ \onlinecite{Matsuda_etal03}).
Thus in the disordered system each chain splits up into
ferromagnetic fragments separated by the holes which induce
antiphase boundaries.
But for the present model the spins form ``helical'' chain fragments
due to the competing intrachain interactions (see Sec.\ \ref{sec_pure}),
and again reverse their direction across a hole.
In diffraction experiments, no indications of such a residual helical ordering
in La$_5$Ca$_9$Cu$_{24}$O$_{41}$ have been found so far.
One should note, however, that hints at incommensurate ordered spin structures,
which could in principle be explained by the presence of a (modified) helical
phase, have been reported
in La$_6$Ca$_8$Cu$_{24}$O$_{41}$.~\cite{Matsuda_etal96}
Note also that in the closely-related spin-chain system Li$_2$CuO$_2$
the helical ordering is destroyed due to the anisotropy and the finite
interchain coupling according to theoretical
calculations.~\cite{Drechsler_etal05}
Thus the existence or non-existence of helical structures appears to be
a delicate question which depends sensitively on details of the interaction
and the lattice geometry (i.e., the coordination number of the interchain
interaction, which is different for La$_5$Ca$_9$Cu$_{24}$O$_{41}$
and Li$_2$CuO$_2$).
\begin{figure}[ht]
  \includegraphics[width=8.6cm]{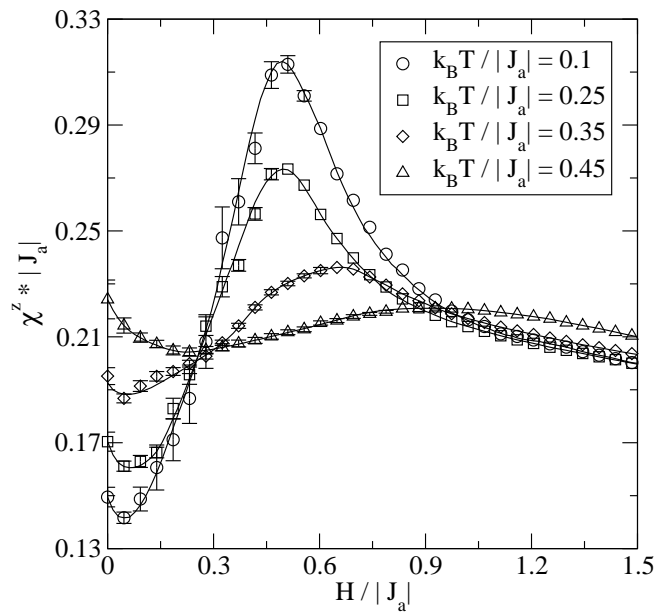}
  \caption
   {
    Susceptibility curves for different temperatures
    (and system size $L=80$).
    As in Fig.\ \ref{fig_chi_T025}, we used a varying number
    of up to 300 disorder realizations (depending on the value of the field)
    to generate the data points.
    The solid lines are guides to the eye.
   }
  \label{fig_chi_multiT}
\end{figure}

Plotting the susceptibility curves for various temperatures
(Fig.\ \ref{fig_chi_multiT}) allows us to draw a more detailed comparison
with the experimental magnetization measurements.
The temperature dependence of both the position and the height
of the spin-flop anomaly resemble the experimental data which will be
presented in Sec.\ \ref{sec_exp} (Fig.\ \ref{fig_chi_multiT_la52}) quite well.

\begin{figure}[ht]
  \includegraphics[width=8.6cm]{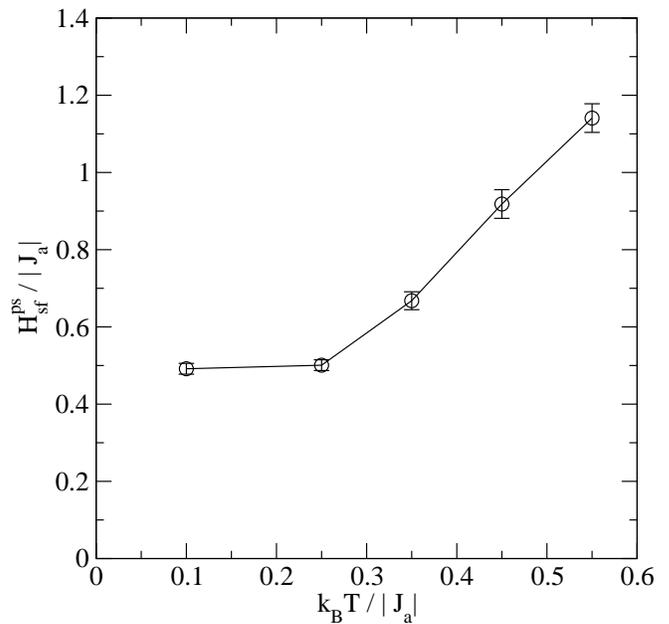}
  \caption
   {
    Magnetic phase diagram of the model with quenched disorder.
    The curve shows the temperature dependence of the ``pseudo'' spin-flop
    field where the susceptibility maximum occurs
    (cf.\ Fig.\ \ref{fig_chi_multiT}).
    All data were obtained using systems of size $L=80$.
   }
  \label{fig_phdiag_disord}
\end{figure}
Finally, we depict the ``magnetic phase diagram'' of our model
in Fig.\ \ref{fig_phdiag_disord}, i.e., the dependence of the ``pseudo''
spin-flop field $H_\mathrm{sf}^\mathrm{ps}$ on temperature.
Again, we find qualitative agreement with the upper line $B_2$
of the experimental phase diagram (Fig.\ \ref{fig_chi_multiT_la52}).
If one converts the theoretical values for $H_\mathrm{sf}^\mathrm{ps}$
into physical units, taking the spin value $S=1/2$ of the Cu$^{2+}$ ions
and the correct $g$-factors \cite{Kataev_etal01} properly into account,
we obtain a ``pseudo'' spin-flop field of approximately 9T
for a temperature of $k_BT/|J_a|=0.1$.
This compares reasonably well with the experimental values.
An even better agreement may be reached by fine-tuning of the interaction
parameters (whose precise values are not known yet), taking into account
quantum effects, or allowing for a (partial) mobility of the holes.

Summarizing, the above findings corroborate the idea that the experimentally
observed broad anomaly in the susceptibility curves can indeed be understood
as a disorder phenomenon due to randomly distributed, quasi-static holes,
which lead to a destruction of the long-range AFM order and, correspondingly,
to a smearing-out of the spin-flop transition.

\section{Experimental results on the ``pseudo spin-flop peak''}\label{sec_exp}

In order to test the numerical predictions of the preceding section,
we here present some of our experimental data on the magnetic properties
of the lightly hole-doped spin chains
in La$_x$(Ca,Sr)$_{14-x}$Cu$_{24}$O$_{41}$.
We studied single crystals of approximately 0.2cm$^3$, grown
by the floating zone technique.~\cite{Ammerahl_etal99a}
For the magnetization measurements a vibrating sample magnetometer (VSM)
was used.
The measurements were performed in magnetic fields up to 16T.
The fields were applied either parallel to the chain direction ($c$ axis)
or perpendicular to the CuO$_4$ plaquettes of the CuO$_2$ chains,
i.e.\ along the easy magnetic axis ($b$ axis).

\begin{figure}[ht]
 \includegraphics[width=8.6cm]{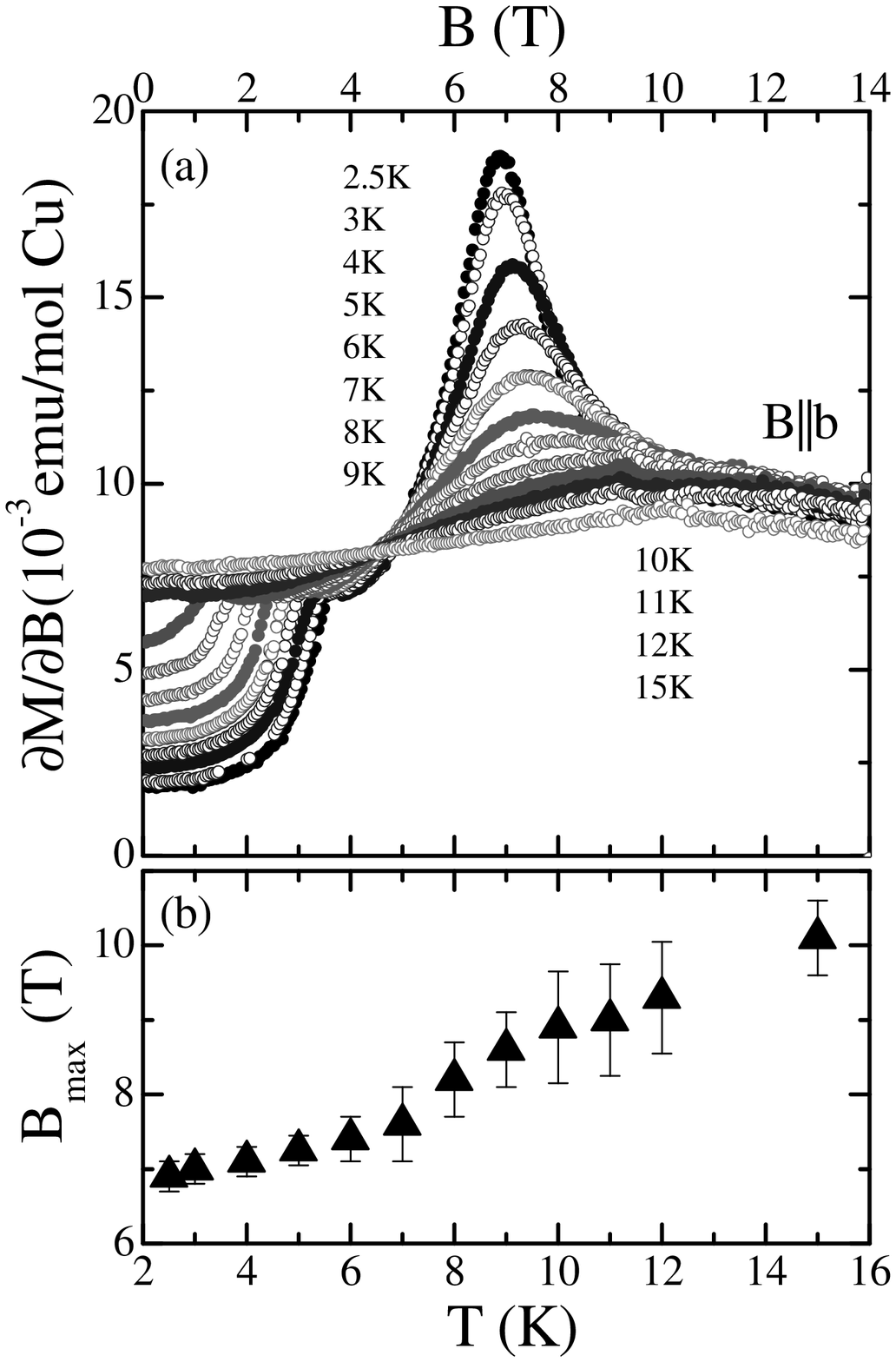}
  \caption
   {
    Susceptibility of La$_{5.2}$Ca$_{8.8}$Cu$_{24}$O$_{41}$
    vs.\ magnetic field $B\|b$ parallel to the easy axis
    for different temperatures, (a), and position
    of the maximum in (a) vs.\ temperature, (b).
   }
  \label{fig_chi_multiT_la52}
\end{figure}
Fig.\ \ref{fig_chi_multiT_la52}a shows the susceptibility
of La$_{5.2}$Ca$_{8.8}$Cu$_{24}$O$_{41}$ vs.\ magnetic field
along the $b$-axis, at different temperatures up to $15\mathrm{K} > T_N =
10.5\mathrm{K}$.
The sharp anomaly at $B_1$, which signals the melting of the long-range
spin order, is visible for all temperatures below $T_N$.
In contrast to the melting of the spin order, the anomaly at $B_2$
is still present for $T>T_N$, where only short-range spin correlations exist.
This fact agrees with the observation that at $T=2.5\mathrm{K}$
the anomaly occurs at fields $B>B_1$, where also only short-range spin order
does exist.
Upon heating, the anomaly is shifted to higher fields.
Moreover, the peak both shrinks and broadens drastically
at higher temperatures.
Comparing the data in Fig.\ \ref{fig_chi_multiT_la52}a with those
in Fig.\ \ref{fig_chi_multiT} illustrates the similarities
between the experimental data and the numerical results.
The broad peak at $B_2$ is well described by the model calculations,
which strongly reinforces the approach presented in Sec.\ \ref{NumSim}.

The temperature dependence of the ''pseudo'' spin-flop transition
in La$_{5.2}$Ca$_{8.8}$Cu$_{24}$O$_{41}$ is summarized
in Fig.\ \ref{fig_chi_multiT_la52}b.
Qualitatively, the presence of the broad anomaly $B_2$ indicates short-range
spin correlations up to 15K.
The temperature dependence of the peak maximum depends roughly linearly
on the temperature, in agreement with the numerical findings
(Fig.\ \ref{fig_phdiag_disord}).
At (7$\pm$1)K, the curvature of $B_2(T)$ slightly changes,
which again resembles the numerical results.

\section{Conclusion and Outlook}

We have presented measurements of the magnetic properties of the anisotropic
spin chains in lightly hole-doped La$_x$(Ca,Sr)$_{14-x}$Cu$_{24}$O$_{41}$,
$x\ge5$.
The experiments suggest that for fields $B\gtrsim4\mathrm{T}$ the system is
characterized by short-range AFM spin order and quasi-static charge disorder.
The susceptibility as a function of the magnetic field $B$ (applied along
the easy axis) shows a broad anomaly instead of a sharp peak,
as one would have expected if the system underwent a spin-flop transition
(and as one indeed observes in the related, but hole-free spin chain compound
Li$_2$CuO$_2$, see Ref.\ \onlinecite{Klingeler_PhD}).
In order to understand these findings theoretically, we have carried out
Monte Carlo simulations of an anisotropic classical Heisenberg model
with quenched holes.
Our numerical data show that the spin-flop transition of the pure model
is smeared out upon introducing quenched holes.
The susceptibility curves at fixed temperature exibit broad peaks and resemble
the experimental data quite well.
At low temperatures, the peak occurs at a field value slightly below
the corresponding spin-flop field of the pure system.
Furthermore, the peak position increases with temperature, similarly as
it is observed in the experiments.
Taken together, our numerical studies corroborate the idea
that the broad anomaly in the experimental susceptibility curves
is essentially a signature of the disorder due to quasi-static holes.

Nonetheless, there remain several challenging questions for future
(experimental and theoretical) work.
One of them concerns the possible mechanisms for the pinning of the holes,
which would explain the occurrence of quasi-static (quenched) charge disorder.
While the destruction of the stripe-ordered phase by an effective,
field-induced attraction of the holes (as proposed
in Ref.\ \onlinecite{Klingeler_PhD}) obviously requires a certain mobility
of the holes, the existence of the broad anomaly in the susceptibility seems
to suggest that pinning might play an important role to understand
the high-field behavior ($B>B_1$).
If one assumes the holes to move freely along the chains, the theoretical
models predict a clustering of the holes upon increasing the field
and no broad anomaly in the susceptibility occurs (see the discussion
in Ref.\ \onlinecite{LS04b}).
Thus one may speculate that as the field becomes large enough ($B>B_1$)
and the holes start to move around, they get trapped at (randomly distributed)
pinning centers and then stay more or less immobile.

Closely related to the pinning of the holes is the possible influence of the
Coulomb interaction, which has been neglected in the theoretical models so far.
The Coulomb repulsion of the holes destabilizes the above-mentioned
clustered structures and would tend to distribute the holes more uniformly
across the system.
It is unclear, however, whether inclusion of the Coulomb interaction
between the holes alone would suffice to predict the existence
of quasi-static disorder for fields $B>B_1$.
It might also be necessary to take the interaction with the La$^{3+}$
and Ca$^{2+}$ ions into account.
Due to their different charges and ionic radii, these might introduce
additional disorder into the system which may turn out to be important
for an understanding of the pinning of the holes.

\acknowledgments

We thank J.\ M.\ Tranquada for a useful discussion.
This work was supported by the Deutsche Forschungsgemeinschaft (DFG)
within SPP 1073 (BU 887/1-3).
W.S., M.H., and R.K.\ gratefully acknowledge financial support by the Deutsche
Forschungsgemeinschaft under Grants No.\ SE 324/4 and KL 1824/1-1,
respectively.

\end{document}